\PassOptionsToPackage{table}{xcolor}
\documentclass[sigconf]{acmart}

\usepackage{soul}
\AtBeginDocument{%
  \providecommand\BibTeX{{%
    \normalfont B\kern-0.5em{\scshape i\kern-0.25em b}\kern-0.8em\TeX}}}

\newcommand{\getcolor}[1]{
    \ifdim#1pt>1pt \cellcolor{green!50}
    \else\ifdim#1pt>0.5pt \cellcolor{green!30}
    \else\ifdim#1pt>0.2pt \cellcolor{green!20}
    \else\ifdim#1pt>0pt \cellcolor{green!05}
    \else\ifdim#1pt<0pt \cellcolor{red!10}
    \else\ifdim#1pt<-0.2pt \cellcolor{red!30}
    \else\ifdim#1pt<-0.3pt \cellcolor{red!50}
    \fi\fi\fi\fi\fi\fi\fi
}

\copyrightyear{2023}
\acmYear{2023}
\setcopyright{acmlicensed}
\acmConference[ARES 2023]{The 18th International Conference on Availability, Reliability and Security}{August 29-September 1, 2023}{Benevento, Italy}
\acmBooktitle{The 18th International Conference on Availability, Reliability and Security (ARES 2023), August 29-September 1, 2023, Benevento, Italy}
\acmPrice{15.00}
\acmDOI{10.1145/3600160.3600193}
\acmISBN{979-8-4007-0772-8/23/08}

\begin{document}

\title{SoK: Modeling Explainability in Security Analytics for Interpretability, Trustworthiness, and Usability}

\author{Dipkamal Bhusal}
%\email{db1702@rit.edu}
\affiliation{%
  \institution{Rochester Institute of Technology}
  %\streetaddress{P.O. Box 1212}
  \city{Rochester}
  \state{NY}
  \country{USA}
}

\author{Rosalyn Shin}
\affiliation{%
  \institution{Independent Researcher}
  %\streetaddress{1 Th{\o}rv{\"a}ld Circle}
  \city{Seoul}
  \country{South Korea}}
%\email{rosalynhjs@gmail.com}

\author{Ajay Ashok Shewale}
\affiliation{%
  \institution{Rochester Institute of Technology}
  %\streetaddress{P.O. Box 1212}
  \city{Rochester}
  \state{NY}
  \country{USA}
}

\author{Monish Kumar Manikya Veerabhadran}
\affiliation{%
  \institution{Rochester Institute of Technology}
  %\streetaddress{P.O. Box 1212}
  \city{Rochester}
  \state{NY}
  \country{USA}
}

\author{Michael Clifford}
\affiliation{%
  \institution{Toyota Motor North America}
  %\streetaddress{8600 Datapoint Drive}
  \city{Mountain View}
  \state{CA}
  \country{USA}}
  %\postcode{78229}}
%\email{michael.clifford@toyota.com}

\author{Sara Rampazzi}
\affiliation{%
  \institution{University of Florida}
  %\streetaddress{1 Th{\o}rv{\"a}ld Circle}
  \city{Gainesville}
  \state{FL}
  \country{USA}}
%\email{srampazzi@ufl.edu}

\author{Nidhi Rastogi}
\affiliation{%
  \institution{Rochester Institute of Technology}
  %\streetaddress{P.O. Box 1212}
  \city{Rochester}
  \state{NY}
  \country{USA}
}
%\email{nxrvse@rit.edu}

\renewcommand{\shortauthors}{Bhusal et al.}

\begin{abstract}
Interpretability, trustworthiness, and usability are key considerations in high-stake security applications, especially when utilizing deep learning models. While these models are known for their high accuracy, they behave as black boxes in which identifying  important features and factors that led to a classification or a prediction is difficult. This can lead to uncertainty and distrust, especially when an incorrect prediction results in severe consequences. Thus, explanation methods aim to provide insights into the inner working of deep learning models. However, most explanation methods provide inconsistent explanations, have low fidelity, and are susceptible to adversarial manipulation, which can reduce model trustworthiness. This paper provides a comprehensive analysis of explainable methods and demonstrates their efficacy in three distinct security applications: anomaly detection using system logs, malware prediction, and detection of adversarial images. Our quantitative and qualitative analysis\footnote{An Institutional Review Board (IRB) approval was taken prior to interviewing experts for qualitative study.} reveals serious limitations and concerns in state-of-the-art explanation methods in all three applications. We show that explanation methods for security applications necessitate distinct characteristics, such as stability, fidelity, robustness, and usability, among others, which we outline as the prerequisites for trustworthy explanation methods.
\end{abstract}






%\thispagestyle{plain}
%\pagestyle{plain}

\settopmatter{printfolios=true}
\maketitle

\section{Introduction}\label{sec:intro}

Deep learning (DL) models have shown high performance in a variety of security applications, %\cite{parra2022interpretable} 
such as security log analysis \cite{deepcase}, vulnerability detection \cite{dam2017automatic}, and malware detection \cite{yerima2018droidfusion}. However, since these models operate by learning complex, non-linear relationships between inputs and outputs, security researchers find it difficult to interpret their inner workings or explain them to other researchers. As a result, deep learning models may not offer insights for their classification decisions, making it difficult to evaluate whether their decision criteria align with domain expert knowledge \cite{alahmadi202299}.

\begin{figure}[h!]
    \centering
 \includegraphics[width=.45\textwidth]{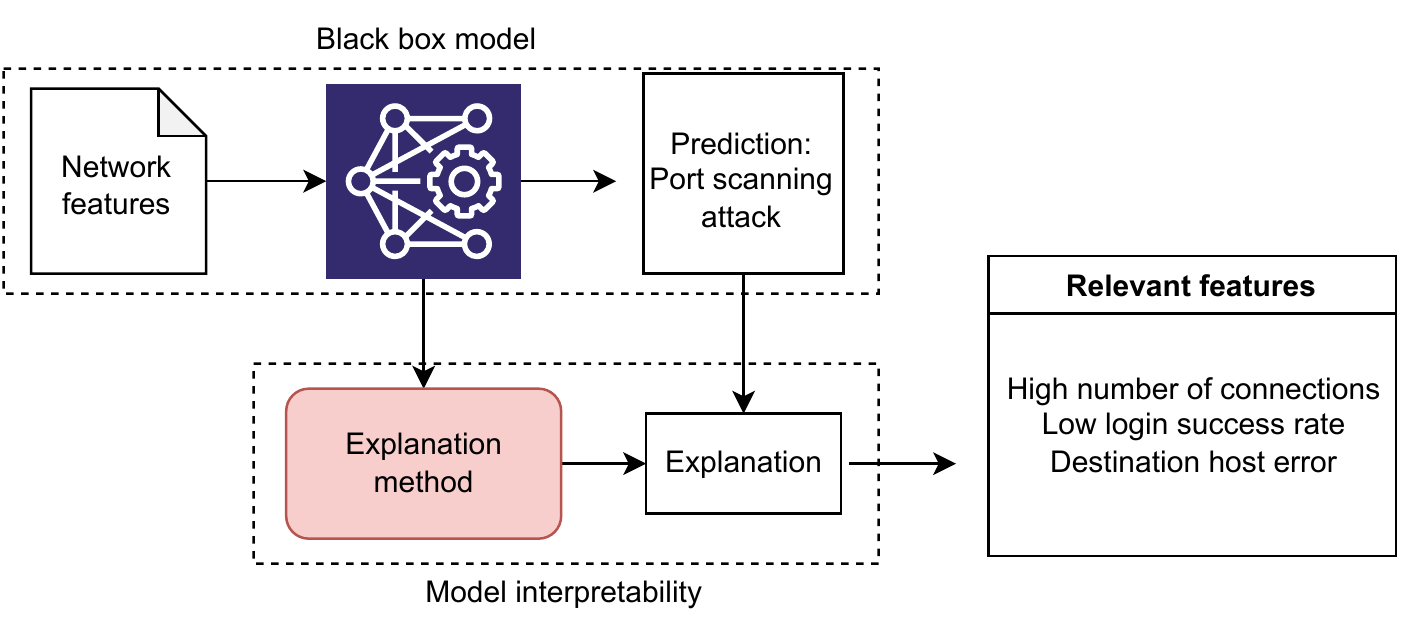} 
    %\caption[caption]{Parsing of unstructured log into structured sequence \footnote{\url{https://github.com/logpai/logparser}}}
    \caption[Caption]{Overview of a post-hoc explanation method in network intrusion detection where the explanation method provides a list of top features relevant for predicting the attack.}
    \label{fig:tree}
\end{figure}

For instance, in the context of intrusion detection, security researchers require insights into the underlying behavior of  intrusion detection systems, which inspect network packet features such as source and destination IP addresses, protocol types, packet lengths, payload contents, and sequence numbers, and then classify packets as suspicious or not. In a \textit{port scanning attack}, model decisions are trustworthy if the model pinpoints the presence of a high number of connections with low duration and low login success rate (see Figure \ref{fig:tree}), reflecting both the structure of the attack and the analyst's domain knowledge for that attack. 

Another challenge resulting from the inherent opacity of deep learning models for security applications is the \textit{ethical concern of bias and discrimination}  in  training data \cite{buolamwini_2018}. In the context of facial recognition systems,  models have been shown to exhibit bias against marginalized groups, particularly people of color, due to a lack of training data diversity. This can lead to false positives or false negatives, as well as perpetuate harmful stereotypes and discriminatory behavior. An explanation method can identify the characteristics, such as facial features, that the deep learning model used to make decisions.  Researchers can then address ethical concerns by adjusting relevant feature weights, and by adding more racially and ethnically diverse data.

\par
Interest in achieving interpretability of DL models has led to research following two main approaches: 1) designing intrinsically interpretable models and 2) using post-hoc explanation methods \cite{murdoch2019definitions}. The latter category involves techniques that analyze model decisions with respect to input features and identify the  features most significant to those model decisions. Examples of these methods include LIME (Local Interpretable Model-agnostic Explanations), SHAP (SHapley Additive exPlanations), Integrated Gradient, PatternNet, and Grad-CAM (Gradient-weighted Class Activation Mapping) \cite{ribeiro2016should,sundararajan2017axiomatic, kindermans2017learning,selvaraju2017grad}. However, not all explanation methods are suitable for implementation in the security domain, which typically utilizes recurrent neural networks (RNNs) and feed-forward neural networks (FNNs) \cite{warnecke2020evaluating}. For example, GradCAM \cite{selvaraju2017grad} is only applicable to convolutional neural networks (CNNs) for images, and PatternNet is limited to FNNs and CNNs. Additionally, there are security and privacy concerns associated with the use of explanation methods in the security domain because of differences in design assumptions, requirements, dataset characteristics, and the model itself \cite{buolamwini_2018, shokri2021privacy}.
% \textcolor{blue}{\textit{rewriting this paragraph as:} Interest in achieving interpretability of deep learning (DL) models has led to research in two main approaches: designing intrinsically interpretable models and using post-hoc explanation methods \cite{murdoch2019definitions}. The latter category involves techniques that analyze the model prediction with respect to input features and identify the most significant features for the model's predictions. Examples methods include LIME (Local Interpretable Model-agnostic Explanations), Integrated Gradient, PatternNet, Grad-CAM (Gradient-weighted Class Activation Mapping) \cite{ribeiro2016should,sundararajan2017axiomatic, kindermans2017learning,selvaraju2017grad}. However, GradCAM \cite{selvaraju2017grad} is only applicable to convolutional neural networks for images, and PatternNet is limited to feed-forward and convolutional networks. Since security models typically consist of recurrent network networks and feed-forward networks, not all explanation methods are suitable for implementation in the security domain \cite{warnecke2020evaluating}.} %Additionally, explanation methods are plagued with security and privacy concerns due to differences in design assumptions, requirements, the nature of the dataset, and the model \cite{vigano2020explainable, nadeem2022sok, shokri2021privacy, capuano2022explainable, guo2018lemna, nyre2021considerations, patel2022model}.}
\par
In this paper, we systematically evaluate the use of state-of-the-art explainability models in security monitoring and provide a critical analysis of their effectiveness. We also review  peer-reviewed articles on  explainable security applications, such as malware detection, network intrusion detection, and adversarial object detection. To ensure we captured the most relevant and recent scientific work in this domain, we employed an approach that involved selecting papers that were published after 2010 and that match at least one of the following criteria: (a) publication in high-ranking journals such as NeurIPS, AAAI, USENIX Security, ACM CCS, SIGSAC, IEEE Security \& Privacy, and ACL, (b) high citation rates, (c) authored by researchers actively publishing on that topic, and (d) availability of open-source code, publicly available libraries. We searched relevant papers across several academic databases, including Google Scholar, IEEE Xplore, and the ACM Digital Library.

\textbf{Main Contributions.} This research makes the following main contributions.
\begin{enumerate}
    \item We perform both quantitative and qualitative analysis to evaluate the efficacy of the most representative explanation methods against distinct security threats. The explanation methods include LIME \cite{ribeiro2016should}, SHAP \cite{lundberg2017unified}, Gradient \cite{simonyan2013deep}, GradientXInput \cite{shrikumar2016not}, Integrated Gradient \cite{sundararajan2017axiomatic}, DeepLIFT \cite{shrikumar2017learning}, GradientShap \cite{erion2021improving},  Occlusion \cite{zeiler2014visualizing}, and DeepAID \cite{han2021deepaid}.

\item We provide a comprehensive analysis of the performance of explanation methods for three distinct security threats. Based on our literature study, we determine that existing research provides inadequate evaluation approaches for a quantitative study.
\item We outline important evaluation methods suitable for security explanations. These methods are covered in detail in Section \ref{sec:evaluation}.
\item %Identify limitations of current explanation methods and highlight concerns with respect to the security domain.
Through qualitative analysis, we determine that existing  explanation methods have very poor usability. To the best of our knowledge, this is the first work to present this form of analysis in security. %Security experts struggled to interpret the explanations presented by different models and often required additional guidance to understand the meaning of the results. This limitation underscores the importance of improving the usability of explanation models to enable effective decision-making. 

%\item We propose a pipeline to address accuracy, privacy, and trust concerns in security applications. 
%Finally, we present our findings, discuss challenges, and propose future research directions to complete our systematic review. %a model pipeline to address the existing challenges of accuracy, privacy, and interpretability. We formalize and propose evaluation methods suitable for security explanation that will help researchers to measure their improvement over other existing works quantitatively.

\end{enumerate}

\textbf{Key Findings.} The following are our key findings from the quantitative and qualitative evaluations using various explanation methods to explain models used for different security use cases.
\begin{enumerate}
    \item  Explanation methods display high disparity in attributing feature relevance, raising reliability concerns (Section \ref{sec:usecaseI}). 
    
    \item Back-propagation-based explanation methods (e.g., Integrated Gradient and DeepLIFT) outperform perturbation methods (e.g., LIME, SHAP) in identifying relevant input features of a model prediction (Section \ref{sec:usescaseII}).
    
    \item The choice of explanation method significantly impacts the distribution of attributed features. While the gradient method is capable of capturing class-discriminative behavior and revealing dispersion in feature distributions of benign and adversarial samples, other methods, such as IntegratedGradient, fail to capture statistically significant dispersion (Section \ref{sec:usecaseIII}).
    
    \item Evaluating the explanation method effectiveness using only quantitative metrics leads to misleading results because these methods can assign high importance to irrelevant features. Conversely, relying exclusively on qualitative explanation evaluations produces incomplete results, as the accuracy, stability, and reliability of the explanations cannot be objectively measured. Consequently, comprehensive evaluations of an explanation method should incorporate both quantitative and qualitative evaluation approaches. However, many explanation methods such as SHAP, GradientShap, DeepLift, Gradient, Integrated Gradient, and Occlusion overlook at least one of these evaluation approaches.
        
    \item It is crucial to ensure that end users can comprehend and understand explainable models. Like DL models, explanation methods are susceptible to adversarial manipulation. An adversary can exploit  explanations to produce random feature attributions or reconstruct significant portions of datasets, compromising privacy. Thus, ensuring that DL model explainability does not compromise reliability is crucial.

    \item Qualitative evaluation of explainability model usability by security experts highlights that explainable methods need to improve both explanation quality and coherence between the explanations and expert knowledge.  Poor coherence may result in explanations that have little applicability to the DL model's decision-making process.

\end{enumerate}

This paper is structured as follows: In Section \ref{sec:explainableAI}, we provide an overview of explainable models, discuss their importance in comprehending black box models, and examine their essential properties. In Section \ref{sec:explainableMethods}, we categorize widely used explanation methods, examine their strengths and weaknesses, and discuss different evaluation criteria in Section \ref{sec:evaluation}. In Section \ref{sec:usecase}, we describe three distinct explanation method applications and examine their performance both qualitatively and quantitatively. In Section \ref{sec:concerns}, we identify major security and privacy concerns related to explanation methods based on our study and experiments. Finally, in Section \ref{sec:challenges}, we present challenges and promising research directions to improve the applicability of explainable AI to security monitoring.
\section{Explainability}\label{sec:explainableAI}

\textbf{Background.} 
Taking inspiration from DARPA's Explainable AI (XAI) program, Vigano et al. proposed \cite{vigano2020explainable}, XSec, a high-level framework, and characteristics of explainable security, challenges, and research opportunities. However, the framework, like \cite{warnecke2020evaluating}, does not analyze methods and solely focuses on providing contexts and example cases. In \cite{hariharan2021explainable}, Hariharan et al. briefly overview explanation methods for four security areas: intrusion detection, malware detection, access control, and threat intelligence, but do not provide guidance on the implementation \cite{hariharan2021explainable}. While Warnecke et al. \cite{warnecke2020evaluating} evaluated explanation methods for deep learning in security and provided quantitative insights for security researchers, qualitative analysis is missing, which may be the preference of some users who want easier-to-understand or more intuitive explanations. In \cite{nadeem2022sok}, Nadeem et al. provide several examples to illustrate the use of explainable methods from both designers' and users' perspectives, providing a comprehensive review of explanation methods. However, they do not evaluate explanation methods, and they provide only one use-case with LIME \cite{ribeiro2016should} and SHAP \cite{lundberg2017unified}, which has been previously deemed unreliable in the prior research \cite{warnecke2020evaluating}.
\par
Measuring a model's accuracy when testing a classification model does not sufficiently describe its performance in real-world applications \cite{doshi2017towards}. The opaque nature of the deep learning model impedes their otherwise impressive performance in diverse fields, making it difficult for end-users to comprehend their functionality and decision-making criteria \cite{rudin2022interpretable}. Therefore, to achieve fine-grained evaluation, %in the explanations, 
gaining deeper insight into the working of the black-box model is important \cite{pieters2011explanation}. Explainability helps to gain such insights, thus promoting trustworthiness and ensuring equity, confidentiality, and dependability \cite{pieters2011explanation}.

\textbf{Definition.} The terms "explainability" and "interpretability" are sometimes used interchangeably; however, they have distinct meanings. Interpretable machine learning involves using constraints to create models easily understood by end-users; for instance, a decision tree \cite{rudin2019stop}. However, achieving high performance with interpretable models can be challenging, particularly for raw data like system logs, where complex deep learning models often outperform simpler models. As a result, monitoring systems for malware classification, vulnerability detection, and anomaly detection have increasingly relied on black-box models to improve performance \cite{berman2019survey}. This reliance on black box models has heightened the need for post-hoc explanation methods that provides explanations for non-interpretable black box models.

%\textbf{Classification. }Explanation methods can be classified into different categories based on various factors, including granularity of explanations (local vs. global explanation), the translucency of application (model agnostic vs. model-specific), type of explanation (post-hoc or interpretable model)\cite{nadeem2022sok}, and method used to compute feature attribution (perturbation vs. gradient-based). However, for security applications, the choice of explanation method is determined based on the specific application. 

\textbf{Classification.} Post-hoc explanation methods can be classified into different categories based on various factors, including granularity of explanations (local vs. global), supported models (model-agnostic vs. model-specific), and type of explanations (feature attribution, rules or counterfactual) \cite{bodria2021benchmarking}. Table \ref{tab:summaryXAI} provides a mapping of various explanation methods to their corresponding class.
\newline
\textit{a. Based on Granularity}: Explanation methods can either be \textit{local} or \textit{global} \cite{murdoch2019definitions}. Global explanations provide an overall understanding of the model's behavior across multiple instances. They identify the most important features that the model uses to make decisions. This approach can help assess model biases and strengths and guide improvements. Local explanations help explain the decision-making process of the model for a specific instance by identifying relevant input features that contribute to the model output.
\newline 
\textit{b. Based on \textit{supported models}:} A model-agnostic explanation method can be used to interpret any type of black-box model (e.g., LIME), whereas a model-specific method can be used to interpret only specific types of networks. For example, GradCAM is used to explain predictions of convolutional neural networks only.
\newline 
\textit{c. Type of explanations:} Post-hoc explanation methods use feature attribution, rules, or counterfactuals to provide explanations on a given test instance. \textit{(a) Feature attribution} assigns a relevance score to each feature, indicating its importance in the model's prediction for the given instance (e.g., LIME \cite{ribeiro2016should}). Feature attribution is also known as importance, relevance, contribution, or scores. Formally, given a black box model $f(.)$ and a test instance $x={x_1, x_2,....x_N}$, where $1...N$ refers to the input features, a feature attribution-based explanation method $\phi (x)$ returns a vector $I_k(x)$ that provides the relevance of the $k$ features. The relevance score can be analyzed to identify the crucial features responsible for a model prediction, making feature attribution-based explanation methods the most popular post-hoc explanation technique \cite{bodria2021benchmarking}. \textit{(b) Rule-based explanations}, commonly used for tabular data, provide a decision rule of the form $x \rightarrow y$, where $x$ represents conditions on input features, and $y$ represents the model prediction (e.g., ANCHOR \cite{ribeiro2018anchors}). \textit{(c) Counterfactual-based explanations} attempt to find the closest instance of opposite prediction, such that the difference in feature distribution of the two samples provides explanations for the model prediction (e.g., DeepAID\cite{han2021deepaid}).

\textbf{Focus of this paper. }In this paper, we evaluate feature attribution-based explanation methods, which can be further classified into perturbation and gradient-based methods. Perturbation-based methods modify the input and obtain feature attribution by either fitting a local interpretable model or observing the change in the corresponding model prediction, as demonstrated by LIME \cite{ribeiro2016should}, SHAP \cite{lundberg2017unified}, and Occlusion \cite{zeiler2014visualizing}. In contrast, gradient-based methods either propagate the model decision to the input layer or utilize the gradient of the output to explain the local decision of neural networks, as exemplified by Gradient \cite{simonyan2013deep} and Integrated Gradient \cite{sundararajan2017axiomatic}.

\textbf{Using explanations. } Understanding a black-box model prediction on test samples is crucial in security, as it can identify potential model weaknesses that attackers can exploit. From the perspective of a system designer, black-box explanations can help improve the security system. For instance, a developer designing a malware classifier can evaluate explanations of the model's predictions on test cases to ensure that it relies on relevant malware features and not on spurious features. From the perspective of a security analyst, explanations can mitigate blind faith in a machine-learning model. Alerts generated by deep learning models with explanation insights can help security analysts make informed decisions about security threats and improve the overall security of the system\cite{alahmadi202299}. 

\textbf{Evaluating explanations. } Explanation methods must satisfy certain properties that guarantee their goodness and usefulness in real-life applications \cite{linardatos2020explainable, nyre2022explainable, robnik2018perturbation}.  These properties include accuracy (capturing relevant features), fidelity (approximating the model prediction), stability (producing consistent results), and certainty (reflecting the certainty of the model). Although many explanation methods focus on the goodness of the explanations \cite{linardatos2020explainable, nyre2022explainable}, the usability and human factors of these methods receive insufficient attention \cite{sokol2020explainability, ehsan2020human}. While quantitative evaluation of explanation methods is crucial, qualitative evaluation is equally important since human end-users will ultimately utilize these explanations. In Section \ref{sec:evaluation}, we provide details on the qualitative and quantitative evaluation of explanation methods.

\section{Explanation Methods}\label{sec:explainableMethods}

We briefly introduce the explanation methods evaluated in this paper. Table \ref{tab:summaryXAI} provides an overview of different methods along with their supported  models, the granularity of explanation, explanation type and supported dataset.

\begin{table} [h!]
%\renewcommand{\arraystretch}{1}
%\small
\caption{Explanation methods \& mapping to their corresponding class. We evaluate methods shown in bold.}
\label{tab:summaryXAI}
\centering
\resizebox{0.45\textwidth}{!}{%
\begin{tabular} {|c|ccc|cc|ccc|ccc|}
\hline
&\multicolumn{3}{c|}{DL Model} &\multicolumn{2}{c|}{Granularity} &\multicolumn{3}{c|}{Explanation} &\multicolumn{3}{c|}{Data Type}\\
\cline{2-4}
\cline{5-6}
\cline{7-9}
\cline{10-12}
Methods & \rotatebox[origin=c]{90} { MLP } & \rotatebox[origin=c]{90} { CNN } & \rotatebox[origin=c]{90} { RNN } &
\rotatebox[origin=c]{90} { Local } &
\rotatebox[origin=c]{90} { Global } &
\rotatebox[origin=c]{90} { Feature attribution } &
\rotatebox[origin=c]{90} { Rules } &
\rotatebox[origin=c]{90} { Counterfactuals } &
\rotatebox[origin=c]{90} { Tabular } &
\rotatebox[origin=c]{90} { Text } &
\rotatebox[origin=c]{90} { Image } \\
\hline

\textbf{Gradient \cite{simonyan2013deep}} & \checkmark & \checkmark & \checkmark & \checkmark & & \checkmark &  & & \checkmark & \checkmark & \checkmark \\ \hline

 \textbf{Occlusion \cite{zeiler2014visualizing}} & \checkmark & \checkmark & \checkmark & \checkmark & & \checkmark & & & \checkmark & \checkmark & \checkmark \\ \hline
LRP \cite{bach2015pixel} & \checkmark & \checkmark & \checkmark & \checkmark & & \checkmark &  & & \checkmark & \checkmark & \checkmark \\ \hline

\textbf{GradientXInput \cite{shrikumar2016not}} & \checkmark & \checkmark & \checkmark & \checkmark & & \checkmark &  & & \checkmark & \checkmark & \checkmark \\ \hline

\textbf{LIME \cite{ribeiro2016should}} & \checkmark & \checkmark & \checkmark & \checkmark & & \checkmark & & & \checkmark & \checkmark & \checkmark\\ \hline
\textbf{DeepLift \cite{shrikumar2017learning}} & \checkmark & \checkmark & \checkmark & \checkmark & & \checkmark &  & & \checkmark & \checkmark & \checkmark \\ \hline

\textbf{Integrated Gradient \cite{sundararajan2017axiomatic}} & \checkmark & \checkmark & \checkmark & \checkmark & & \checkmark &  & & \checkmark & \checkmark & \checkmark \\ \hline

\textbf{SHAP \cite{lundberg2017unified}}  & \checkmark & \checkmark & \checkmark & \checkmark & & \checkmark & & & \checkmark & \checkmark & \checkmark \\ \hline
PatternNet \cite{kindermans2017learning} & \checkmark & \checkmark & & \checkmark &   &  \checkmark &  & & \checkmark & \checkmark & \checkmark\\ \hline
GradCAM \cite{selvaraju2017grad} & \checkmark & \checkmark & & \checkmark &   &  \checkmark &  & &  &  & \checkmark\\ \hline
Smoothgrad\cite{smilkov2017smoothgrad} & \checkmark & \checkmark & \checkmark & \checkmark & & \checkmark &  & & \checkmark & \checkmark & \checkmark \\ \hline

ANCHOR \cite{ribeiro2018anchors} & \checkmark & \checkmark & \checkmark & \checkmark & & & \checkmark & & \checkmark &  &  \\ \hline
LEMNA \cite{guo2018lemna} & \checkmark & \checkmark & \checkmark & \checkmark & & \checkmark & & & \checkmark & \checkmark & \checkmark\\ \hline
%RTIS\cite{chen2019looks} & \checkmark & \checkmark & & \checkmark &   &  \checkmark &  & &  &  & \checkmark\\ \hline
ConceptShap \cite{yeh2020completeness} & \checkmark & \checkmark & \checkmark & &  \checkmark & \checkmark & &  & & & \checkmark  \\ \hline
%COIN \cite{coin} & \checkmark & \checkmark & & \checkmark &   & \checkmark &   & & \checkmark &   & \\ \hline
\textbf{GradientShap\cite{erion2021improving}} & \checkmark & \checkmark & \checkmark & \checkmark & & \checkmark &  & & \checkmark & \checkmark & \checkmark \\ \hline

Noisegrad \cite{bykov2022noisegrad} & \checkmark & \checkmark & \checkmark & \checkmark & & \checkmark &  & & \checkmark & \checkmark & \checkmark \\ \hline

\textbf{DeepAid  \cite{han2021deepaid}}  & \checkmark & \checkmark & \checkmark & \checkmark &   & & & \checkmark & \checkmark & \checkmark & \checkmark \\ \hline

%Protonet [28] & \checkmark & \checkmark & \checkmark & &  \checkmark &  & &\checkmark  & & & \checkmark  \\ \hline

\end{tabular} %
}
\end{table}

\textbf{LIME and LEMNA \cite{ribeiro2016should,guo2018lemna}:} LIME (Local Interpretable Model-agnostic Explanation) and LEMNA (Local
Explanation Methods using Nonlinear Approximation) are techniques used to generate local explanations for black box models. LIME utilizes linear regression to generate an interpretable surrogate model by perturbing a test instance and creating new data samples. The weights of the surrogate model now explain predictions, where higher positive weights mean high attributing features. However, LIME's explanations can also be unreliable and manipulable through careful data manipulation \cite{slack2020fooling}. LIME's assumption of feature independence and lack of feature combination can limit its applicability in practical applications. LEMNA addresses this limitation using a mixture regression model with a fused lasso penalty to capture feature dependency. However, recent studies have shown that LEMNA may not perform as well as LIME and SHAP in security applications\cite{warnecke2020evaluating}.

\textbf{SHAP \cite{lundberg2017unified}:} SHAP (Shapley Additive Explanation), like LIME, generates local explanations for black box models. However, it takes a game-theoretic approach to compute the Shapley values for each feature and then uses them for feature attribution. SHAP introduces fundamental properties for feature attribution and shows that Shapley values only satisfy these properties. Like LIME, SHAP generates new samples around a given sample, obtains predictions from the black box model, and uses the dataset to fit an interpretable linear model. However, unlike LIME, SHAP weights the new instances according to the weight a coalition would receive in the Shapley value estimation rather than their closeness to the original sample. While SHAP is computationally slower than LIME, it offers more robust explanations by incorporating game-theoretic principles but is susceptible to data manipulations through careful data manipulations comprising the reliability of the explanations\cite{slack2020fooling}.

%\textbf{Local Explanation Method using Nonlinear Approximation (LEMNA) \cite{guo2018lemna}:} LEMNA is an explanation method designed explicitly for security applications. It approximates a local boundary with a simple interpretable model, similar to LIME, but addresses feature dependency and non-linear local boundaries. LEMNA utilizes a mixture regression model with a fused lasso penalty to capture feature dependency. However, recent studies have demonstrated that LEMNA performs worse than LIME or SHAP, even in security applications \cite{warnecke2020evaluating}.

%\textbf{Contextual Outlier Interpretation (COIN) \cite{coin}:} COIN approaches the task of interpretation as a classification task by partitioning data around each outlier and training a linear support vector machine (SVM) between anomalies and normal data. The model parameters from surrogate SVMs are then used to provide feature attribution for the explanation. COIN is not suitable for high-dimensional data \cite{yang2021cade}.

\textbf{Occlusion \cite{zeiler2014visualizing}:} Occlusion is a perturbation-based explanation method that replaces the input features with baselines and computes the difference in output. The user can provide baseline values depending on the use case, and the corresponding output difference provides the feature attribution.

\textbf{Gradient \cite{simonyan2013deep}:} Gradient, also known as saliency, computes the gradient of the class score with respect to the input. This gradient measures the change in output predictions given the change in input features. This score is used as feature attribution. A simple improvement over this method was proposed in \cite{shrikumar2016not}, called GradientXInput, which multiplies the gradient with input features.

\textbf{Integrated gradient (IG) \cite{sundararajan2017axiomatic}:} Gradient simply calculates the gradient of the output with respect to the input feature, but even if a network relies heavily on a particular feature, the gradient of the class score concerning that feature may have small magnitudes. This issue can occur in deep neural networks (DNNs) due to saturation in the training process. To address this problem, the integrated gradient method accumulates gradients along a linear path from the baseline to the given test sample rather than using simple gradients. The selection of the baseline depends on the specific use-case \cite{sturmfels2020visualizing}.

% SmoothGrad \cite{smilkov2017smoothgrad} and NoiseGrad \cite{bykov2022noisegrad} are improvements over gradient based explanation methods. For a given test sample, Smoothgrad adds gaussian noise to generate $n$ number of samples, and their feature attributions are averaged to compute the final attribution. Instead of adding noise to the input, NoiseGrad introduces noise to the model parameters, generate $n$ models, and takes an ensemble of explanations. 
GradientSHAP \cite{erion2021improving} combines ideas from SHAP \cite{lundberg2017unified} and SmoothGrad \cite{smilkov2017smoothgrad} with integrated gradient \cite{sundararajan2017axiomatic}. Instead of picking one baseline, it randomly selects a baseline from a distribution of baselines, most commonly from the training set, computes attributions, and averages the result.

\textbf{DeepLIFT \cite{shrikumar2017learning}:} DeepLIFT computes the feature attribution score by comparing the activation of each neuron to a `reference activation.' The difference between the two contributes to the feature attribution score. Similar to integrated gradient, the reference activation is selected based on the specific problem at hand, which is critical for obtaining accurate feature attribution. This method is equivalent to layer-wise relevance propagation (LRP) \cite{bach2015pixel}. %DeepLiftSHAP extends the existing DeepLift algorithm to approximate SHAP values \cite{lundberg2017unified}. DeepLift method is equivalent to LRP \cite{bach2015pixel}.

%\textbf{Class activation maps}: Grad-CAM and Grad-CAM++ are types of class activation maps that compute feature-importance maps of convolutional layers with respect to particular classes of model prediction. These Methods backpropagate  the gradient information on a model prediction to the convolution layer and obtain the coarse localization map of feature attribution. These are suitable only for CNN networks.

%\textbf{Layer-wise relevance propagation (LRP) \cite{bach2015pixel}:} LRP proposes layer-wise relevance propagation to distribute relevance scores from the output layer to the input layer (features) and generate explanations. LRP operates under the assumption that classifiers can be decomposed into several layers of computation ($l$) and that the sum of relevance scores contributed by all neurons at each layer ($R_d^l$) is equal. Studies by \cite{shrikumar2017learning} have demonstrated that this method is equivalent to DeepLIFT.
 
%\textcolor{red}{there is an error here, rendering both x as same in the pdf}
\textbf{DeepAID \cite{han2021deepaid}:} DeepAID proposes an explanation method for security applications using optimization based on specific security constraints. This method finds a reference normal sample $x*$ for a given test anomaly $x$ such that the difference between $x*$ and $x$ provides the explanation.

%Backpropagation-based methods are typically more stable than perturbation-based explanation methods. However, they can also assist an adversary in better-estimating gradients in black-box attacks \cite{quan2022amplification} and reconstructing the underlying model \cite{milli2019model}. Additionally, these methods can be manipulated with careful data manipulations, rendering them unreliable \cite{ghorbani2019interpretation} \cite{zhang2020interpretable}.

\textbf{Limitations:} Despite the availability of a wide range of explanation methods for gaining insight into model predictions, there are inherent issues with existing feature attribution methods. Many explanation methods exhibit class-invariant behavior, producing similar feature attribution regardless of the predicted class \cite{nielsen2022robust}. Some methods have also been shown to be insensitive to model parameters; even when random numbers replace the parameters of a trained neural network, the feature attribution does not change significantly \cite{adebayo2018sanity}. Furthermore, explanation methods can produce highly unstable explanations \cite{alvarez2018robustness} and are susceptible to adversarial manipulations leading to unreliable results \cite{ghorbani2019interpretation, heo2019fooling, slack2020fooling, zhang2020interpretable}. Gradient-based explanation methods, in particular, can help an adversary better estimate gradients in black-box attacks \cite{quan2022amplification} and even reconstruct the underlying model \cite{milli2019model}.
\vspace{-0.15cm}
\section{Evaluation of explanation methods}\label{sec:evaluation}
We present comprehensive testing and evaluation metrics to ensure that explanations are of high quality and practical value consistent, and meet consensus. Drawing on our extensive study, we outline critical evaluation methods suitable for security explanations. First, we provide three major categories to evaluate explainability methods \cite{doshi2017towards}, a functionally-grounded explanation as one group and application- and human-grounded as another group.
\subsection{Functionally-grounded} 
To assess the efficacy of a proposed explanation method, we need to employ quantitative metrics that utilize formal definitions and properties of explanation quality. This evaluation approach does not necessitate human validation and instead relies entirely on the definitions of relevant features of explanation methods and their mathematical representations. The following are the functionally-grounded evaluation criteria: 

%\textbf{Definition:} Let us consider $f(x)$ represents a black box model trained to perform a security task (for example: malware detection or log analysis). We employ a post-hoc explanation method $I(.)$ to obtain a set of important features $I_k(x')$ where $k$ is the number of features extracted as important and $x'=\{x'_1, x'_2, x'_3,.....x'_N\}$ is the test sample under evaluation with $N$ dimensions and label $t$. 

 \textbf{1. Faithfulness:} It measures the accuracy of an explanation method in capturing relevant features for a given test sample. It is evaluated by computing the correlation between feature attribution and probability drops when relevant features are modified \cite{alvarez2018towards}.

\textbf{2. Monotonicity:} It measures the attribution faithfulness by evaluating if incrementally adding important features improves the model performance \cite{arya2019one}. A monotonic increase in performance as more features are added indicates that the explanation method captured the relevant features.

\textbf{3. Continuity by Local Lipschitz Estimate:} It measures the coherence of the explanation method for similar test inputs. Called \textit{explanation continuity}, this metric is evaluated using the Lipschitz constant \cite{alvarez2018towards} to compare the explanations generated for a given test input and its neighbor samples in a neighborhood of size $\epsilon$.

\textbf{4. Max-Sensitivity:} It approximates change of explanation under slight perturbation using Monte Carlo sampling \cite{bhattevaluating}. A lower max-sensitivity indicates a better explanation method. %The lower the max-sensitivity, the better an explanation method is.

\textbf{5. Relative output stability:} It measures the stability of an explanation with respect to changes in the output logits of the model \cite{agarwal2022rethinking}.

\textbf{6. Sparsity:} It evaluates the feasibility of an explanation by measuring the number of features deemed relevant. %This is because the human brain can process a limited number of features simultaneously. 
It is computed using the Gini Index applied to the vector of the absolute values of attributions  \cite{chalasani2020concise}.

\textbf{7. Complexity:} It measures the entropy of the fractional contribution of any $i^{th}$ feature to the total magnitude of the attribution \cite{bhattevaluating}. It uses all features of the given test sample to explain a model prediction.

\textbf{8. Model parameter randomization :} It measures the difference in feature attributions produced by an explanation method, provided that the model parameters are randomly modified. The difference is computed as the correlation between the original feature attribution and the new attribution \cite{adebayo2018sanity} 

\subsection{Application- and Human-grounded} 
A formal evaluation process should validate an explanation method with real-world applications. And the most effective way involves designing experiments and evaluation by domain experts since the expert can confirm the usability of the explanation method in assisting them with their tasks. For security-related applications, we propose to utilize the following evaluation criteria and measure the usability of an explanation method:
\textbf{1. Expertise Level:} It measures the expertise required to effectively comprehend and utilize explanation methods. Suppose an explanation contains numerous security domain-specific jargon or complex logic that must be understood to comprehend the decision explanations. In that case, the method will necessitate a high level of expertise (e.g., Level 5). Conversely, if the method provides simple explanations (e.g., "Malware x should be deleted because it will leak what you type in a web browser"), it requires little or no expert knowledge, making it Level 1.

\textbf{2. Explanation Type:} For textual or tabular datasets, the explanation typically involves ranking the input features according to the importance of the explanation method in making its prediction rather than visual representations, as in image datasets.

\textbf{3. Coherence \cite{miller2017explainable}:} It evaluates the consistency of the explanation methods with the end-users domain knowledge. An end-user of an explanation method may possess prior knowledge about the application and data, allowing them to assess the validity of the explanation output and if they align with the expert's experience.

\textbf{4. Actionability \cite{sokol2020explainability}:} It measures the usability of explanations with respect to a given application, as an end-user typically prefers explanations that can be applied as guidelines towards a final decision \cite{krause2016interacting}. The simplicity of the explanation is particularly relevant, as it should be concise and to the point such that no further inquiry into the problem is required.
\section{Use Cases}\label{sec:usecase}

%\st{Post-hoc explanation methods are employed to comprehend the prediction of a black-box model on test cases. This is advantageous for end-users seeking confidence in the model's predictions or troubleshooting the model. Explanation methods can also help in detecting adversarial samples.} 
In this section, we present three use cases, each of which corresponds to a distinct security threat:  network log anomaly detection, malware detection, and adversarial image object detection. For each use case, we conduct a quantitative analysis using a deep learning model trained on a dataset representative of that specific threat. We also analyze several explanation methods suitable for that data type. Multiple explanation metrics are used to evaluate the explanation methods using Captum \cite{kokhlikyan2020captum} and Quantus \cite{hedstrom2023quantus}.
% \st{We employ Captum \cite{kokhlikyan2020captum} to generate explanations and Quantus \cite{hedstrom2023quantus} to evaluate quantitative metrics.}
To complement the quantitative analysis, we also conduct qualitative analysis by interviewing security experts specializing in one or more threats addressed by the three use cases. We pose questions covering the advantages, benefits, and challenges associated with using explanation methods in solving domain-specific problems. The questions align with the four criteria discussed in section \ref{sec:explainableMethods}. Code is available\footnote{\url{https://tinyurl.com/3dcbnxb4}}.

\subsection{Network Log Anomaly Detection}\label{sec:usecaseI}
\textit{A Security Operations Center (SOC) analyst receives an alert from a security monitoring tool that analyzes system logs. To investigate this alert, the analyst utilizes the explanation tool to analyze the sequence of log events and validate the alert.}

\textit{\ul{Model \& Dataset}}: For log anomaly detection, we opted to use the widely used DeepLog architecture \cite{deeplog} and Hadoop Distributed File System (HDFS) \cite{xu2009detecting} dataset. The HDFS dataset comprises of logs generated from running map-reduce jobs on 200 Amazon EC2 nodes and comprises 11.2 million log entries, with 2.9\% labeled as anomalies. The log entries are created from 29 distinct log events, each mapped to a unique log key and arranged in a sequence. We trained a Long Short-Term Memory (LSTM) sequence model \cite{deeplogGit} with a window size of 10. This implies that a history of 10 event sequences in the logs is required to predict the next event.

\begin{table}[!ht]
\caption{\small Explanation for malicious event detection in security logs using various explanation methods, with positive attributions highlighted in green and negative attributions highlighted in red. The intensity of the color signifies the relative magnitude of the attribution score.}\label{tab:usecaselog}
\begin{center}
\resizebox{0.48\textwidth}{!}{
\begin{tabular}{cccccccc}
\hline
Event description & Gradient & GradientXInput & IG & DeepLift & LIME & SHAP & Occlusion \\
\hline
 Receiving blk* src\&dest:*  &\getcolor{1.1080} 4 & \getcolor{1.1080} 4 & \getcolor{-0.6700} 4 & \getcolor{1.1080} 4 & \getcolor{-0.5240} 4 & \getcolor{-0.0080} 4 & \getcolor{-0.5320} 4 \\

PktResponder* for blk* terminating &\getcolor{0.4370} 10 & \getcolor{0.4370} 10 & \cellcolor{green!50} 10 & \getcolor{0.4370} 10 & \cellcolor{green!50} 10 & \cellcolor{green!50} 10 & \cellcolor{green!50} 10 \\

 PktResponder* Exception &\getcolor{0.1970} 9 &\cellcolor{red!50} 9 & \getcolor{0.0840} 9 & \cellcolor{red!50} 9 & \getcolor{0.0000} 9 & \cellcolor{green!30} 9 & \getcolor{-0.1190} 9 \\

Exception in receiveBlock for blk & \getcolor{0.0610} 13 & \getcolor{-0.0610} 13 & \getcolor{-0.0040} 13 & \getcolor{-0.0610} 13 & \getcolor{0.0000} 13 & \getcolor{0.0070} 13 & \getcolor{-0.0790} 13 \\

writeBlock* received exception &\getcolor{0.1230} 6 & \getcolor{0.0520} 6 & \getcolor{0.1230} 6 & \getcolor{0.1183} 6 & \getcolor{-0.0084} 6 & \getcolor{0.0037} 6 & \getcolor{0.0529} 6 \\

PktResponder* for blk* Interrupted &\getcolor{-0.0182} 7 & \getcolor{-0.0974} 7 & \getcolor{-0.0182} 7 & \getcolor{-0.0068} 7 & \getcolor{0.0000} 7 & \getcolor{0.0065} 7 & \getcolor{-0.1274} 7 \\

PktResponder* for blk* terminating & \getcolor{0.4224} 10 & \getcolor{0.1667} 10 & \getcolor{0.4224} 10 & \getcolor{0.4612} 10 & \cellcolor{green!30} 10 & \getcolor{-0.0002} 10 & \cellcolor{green!30} 10 \\

 Exception in receiveBlock for blk &\cellcolor{red!50} 13 & \getcolor{-0.0438} 13 & \getcolor{-0.0423} 13 & \getcolor{-0.0632} 13 & \getcolor{0.0000} 13 & \cellcolor{red!50} 13 & \getcolor{-0.0465} 13 \\

writeBlock* received exception  & \getcolor{0.0293} 6 & \getcolor{0.0076} 6 & \getcolor{0.0293} 6 & \getcolor{0.0851} 6 & \getcolor{0.0000} 6 & \getcolor{-0.0079} 6 & \getcolor{-0.0001} 6 \\

PktResponder* for blk* terminating & \getcolor{0.0482} 10 & \getcolor{0.0409} 10 & \getcolor{0.0482} 10 & \getcolor{0.4431} 10 & \getcolor{0.0000} 10 & \getcolor{-0.0079} 10 & \getcolor{0.0287} 10 \\

\hline
\end{tabular}}
\end{center}
\end{table}

We instantiated the network, trained it on the HDFS training dataset, and saved the final trained model for inference. We then assessed the model's performance in anomaly detection by testing it on normal and anomalous test datasets. Evaluation exhibits high performance for system anomaly detection with a precision score of 95.28\%, recall score of 93.37\%, and f1-score of 94.32\%.

\begin{table*}[!ht]
\centering 
\caption{Explanation for the malicious event detection in security logs using DeepAID. The difference column displays which event id was replaced to produce a benign sequence from an anomaly.}
\label{tab:usecasedeepaid}
\resizebox{0.7\textwidth}{!}{%
\begin{tabular}{ccccc}
\toprule
Anomaly Event ID & Event description & Diff & Benign Event ID & Event description \\
\midrule
 4 &      Receiving blk* src\&dest:*      &       &  4 &      Receiving blk* src\&dest:* \\      
 
 10  &  PktResponder* for blk* terminating &      & 10  &  PktResponder* for blk* terminating \\
 
9   &      PktResponder* Exception      &       &  9   &       PktResponder* Exception     \\ 
13  &  Exception in receiveBlock for blk* &       &  13  &  Exception in receiveBlock for blk* \\ 
6   &  writeBlock* received exception*   &      &  6  &   writeBlock* received exception* \\ 
7  & PktResponder* for blk* Interrupted. &      &  7   & PktResponder* for blk* Interrupted \\ 
 10  &  PktResponder* for blk* terminating &      & 10  &  PktResponder* for blk* terminating \\
 13  &  Exception in receiveBlock for blk* &       &  13  &  Exception in receiveBlock for blk* \\ 
 6   &  writeBlock* received exception*   &      &  6  &   writeBlock* received exception* \\ 
 
  10  &  PktResponder* for blk* terminating &      & 10  &  PktResponder* for blk* terminating \\
9   &      PktResponder* Exception      &   !=    &  1   &       Verification succeeded     \\ 

\bottomrule
\end{tabular}%
}
%\label{tab1}
\end{table*}
\textit{\ul{Evaluating Explanation Methods}}: We perform a test case analysis to evaluate different explanation methods' attribution of the input sequence. Since the input should be an anomalous sequence, we choose the following sequence classified as anomalous by DeepLog, [4, 10,  9, 13,  6,  7, 10, 13,  6, 10]. Next, we apply various explanation methods to compute the attribution weight of each event in the test sequence. Table \ref{tab:usecaselog} shows the attribution weights assigned to log events in the sequence, with positive weights highlighted in green. The darker the color, the higher the weight. The red color signifies a negative influence on determining the anomalous event.

\hspace{-3mm}\textit{(a) Quantitative Analysis}: Although \textsf{Gradient, InputXGrad}, and \textsf{DeepLift} assign the highest relevance to the first event ID-4 in the sequence, other methods suggest negative relevance for event ID-4 in relation to the anomalous prediction. Additionally, \textsf{LIME} assigns zero weight to certain event IDs, implying that they had no relevance to the model prediction. Integrated Gradient (IG), LIME, and Occlusion assign the highest attribution to event ID-10. On the other hand, \textsf{DeepAID} (as shown in Table \ref{tab:usecasedeepaid}) follows a different approach; it identifies the closest normal instance to the anomaly and replaces the event ID corresponding to the anomaly with a benign log event. In the given test sequence, \textsf{DeepAID} highlights event ID-9 as the point where an anomaly occurs in the sequence.
\par
A SOC analyst can use the information presented in Table \ref{tab:usecaselog} in evaluating a model-generated alarm. The explanation methods are used to examine prior events that are attributed as highly relevant. Alternatively, using DeepAID, the analyst can inspect the single target log event and directly observe the decision criteria of the deep learning model instead of blindly relying on it. On the flip side, we observe \textit{inconsistencies} in the explanations provided by existing methods, which raises \textit{concerns about their reliability}. Therefore, we conduct qualitative analysis to validate the findings and assess the quality of the explanations.

\begin{table}[h!]
\centering
\caption{Qualitative Evaluation of the 3 Security Applications}
\label{tab:qualeval}
\small
\resizebox{0.45\textwidth}{!}{%
\begin{tabular}{@{}cccc@{}}
\toprule
    & \textbf{Network Log Anomaly} & \textbf{Malware}  & \textbf{Adversarial Image} \\ \midrule
\textbf{Level of Expertise}   & 2 & 2 & 1, 2 \\ \hline
\textbf{Explanation Type} & Textual & Tabular & Visual \\ \hline
\textbf{Coherence}   & Low & Medium & Low \\ \hline
\textbf{Actionability} & Low & Low-Medium & Low  \\ \bottomrule

\end{tabular}%
}
\end{table}

\textit{(b) Qualitative Analysis}: We interviewed a security expert with over 10 years of experience who is well-versed in network log anomaly detection and the HDFS dataset but not familiar with explainability methods. The expert was 
asked to interpret the Tables \ref{tab:usecaselog}, \ref{tab:usecasedeepaid}. The expert needed guidance in navigating the information presented in tables, especially the meaning of the colors in Table \ref{tab:usecaselog}. After a brief introduction, the expert began analyzing the explanations based on their prior knowledge and experience, demonstrating \textit{coherence}. The expert expressed surprise over the choice of event ID-4 as dark green (a very important feature) by half the models. According to the expert, event ID-10, not 4, should have been the top choice, and 4 should not have been chosen at all. The expert admitted that the result could vary for different datasets but still believed that 4 should not be a top choice. The expert \textit{wished} to see confidence levels of %or accuracy 
rankings among the methods, without which it was difficult to choose which model(s) to use for decision-making purposes.

\subsection{Malware Classification}\label{sec:usescaseII}
\textit{A security researcher designs a PDF malware classifier and wants to investigate if the black box model utilizes relevant properties of malware PDFs in making accurate predictions.}

\textit{\ul{Model \& Dataset}}: 
We trained a classifier to perform PDF malware detection, utilizing the Mimicus dataset \cite{guo2018lemna}, which is specifically designed for malicious PDF detection. It includes 135 features such as document structures, counts of JavaScript, counts of JS objects, numbers of sections, and fonts presented in a tabular format. To build the PDF malware classifier, we trained a 3-layer neural network similar to \cite{warnecke2020evaluating} and evaluated the performance on the test set. The model achieved an accuracy of 0.996, a precision of 0.994, and a recall of 0.997.

\textit{\ul{Evaluating Explanation Methods}}: 
We utilized several explanation methods to identify the top 10 relevant features for a given set of malware PDFs in our testing phase. Table \ref{tab:pdfUseCaseExplanations} summarizes the relevant features identified by each explanation method, ordered by their level of importance.

\hspace{-3mm}\textit{(a) Quantitative Analysis}: Most methods, except for Occlusion, successfully capture key malicious features in PDF malware files such as $count\_js$, $pos\_eof\_min$, and $count\_javascript$ \cite{laskov2014practical}. However, the order of these features varies considerably across the different methods. GradientXInput, Integrated Gradient, DeepLIFT, and GradientShap are gradient-based approaches with similar attribution rankings. Most methods also assign relevance to non-indicative features of maliciousness, such as $keywords\_num$, $version$, and $count\_box\_letter$. In contrast, perturbation-based methods such as LIME, SHAP, and Occlusion showed very different feature rankings.
\par
Although feature rankings are insightful, they do not necessarily align with domain expert knowledge of PDF malware. To assist end-users without expertise in explanations, an explanation method should provide actionable guidelines on effectively utilizing the explanations.

\textit{(b) Qualitative Analysis}: We interviewed a security expert with over 10 years of experience. They are well-versed in building defense systems against malware attacks. The expert had used explanation methods once before, therefore, was familiar with explainability methods. The expert was asked to interpret Table \ref{tab:pdfUseCaseExplanations} and immediately reported that they would only use the models with `pos\_image\_avg' as the top feature, followed by `pos\_image\_max' and `pos\_image\_min' (e.g., IG, DeepLift). If these features were not present, they would discard the models. The expert found the features identified by SHAP interesting (e.g., `author\_dot') as they had not previously considered it an indicator of malware. However, the expert noted the need for further explanation on why certain features were chosen as top predictors, such as actual examples where the feature was helpful and a verbal explanation. %\textcolor{blue}{confidence in the rankings}
% \textcolor{orange}{1) Explanation Quality and Type: tabular; low-medium\\
% 2) Level of Expertise: Level 2\\
% 3) Coherence: medium\\
% 4) Actionability: low-medium
% }

% \usepackage{booktabs}
% \usepackage{graphicx}
\begin{table*}[]
\centering
\caption{Relevant features extracted by each explanation method sorted in decreasing order of importance.}
\label{tab:pdfUseCaseExplanations}
\resizebox{\textwidth}{!}{%
\begin{tabular}{@{}llllllll@{}}
\toprule
\textbf{Gradient} & \textbf{GradientXInput} & \textbf{Integrated Gradient} & \textbf{DeepLift} & \textbf{GradientShap} & \textbf{LIME} & \textbf{SHAP} & \textbf{Occlusion} \\ \midrule
count\_stream\_diff & pos\_eof\_min & pos\_image\_avg & pos\_image\_avg & pos\_box\_max & count\_js & pos\_image\_max & createdate\_tz \\
count\_js & pos\_image\_avg & pos\_image\_max & pos\_image\_max & pdfid1\_len & len\_obj\_min & pos\_eof\_min & createdate\_ts \\
len\_obj\_min & pos\_image\_max & pos\_image\_min & pos\_image\_min & pdfid0\_len & keywords\_uc & pos\_image\_min & count\_startxref \\
count\_javascript & pos\_image\_min & pos\_eof\_min & pos\_eof\_min & pos\_eof\_min & pdfid1\_oth & pos\_image\_avg & count\_box\_letter \\
len\_stream\_avg & len\_obj\_min & len\_obj\_min & len\_obj\_min & pdfid1\_num & pdfid0\_oth & author\_dot & count\_page\_obs \\
ratio\_size\_obj & pos\_eof\_avg & pos\_eof\_avg & pos\_eof\_avg & pdfid\_mismatch & pos\_acroform\_avg & len\_obj\_min & createdate\_version\_ratio \\
producer\_uc & pos\_box\_min & createdate\_tz & createdate\_tz & pos\_image\_max & keywords\_lc & pos\_eof\_avg & moddate\_mismatch \\
keywords\_num & pos\_page\_min & pos\_box\_min & pos\_box\_min & producer\_uc & count\_js\_obs & count\_endstream & count\_box\_other \\
len\_obj\_avg & version & version & version & pos\_page\_max & count\_action & pdfid1\_oth & createdate\_mismatch \\ 
count\_box\_a4 & author\_uc & moddate\_tz & moddate\_tz & pos\_image\_avg & count\_javascript & pdfid0\_oth & count\_page \\ \midrule
\end{tabular}%
}
\end{table*}

 %We can see a significant difference in SHAP explanation with LIME and LEMNA. While LIME and LEMNA share 40\% of relevant features, only a few SHAP features overlap with the other two methods. H

\subsection{Adversarial Image Detection}\label{sec:usecaseIII}
\textit{A security researcher detecting threats to autonomous systems utilizes explanation methods to comprehend model predictions and explore adversarial image detection.}

\begin{figure}[h!]
    \centering
    \resizebox{.6\columnwidth}{!}{
 \includegraphics[width=.4\textwidth]{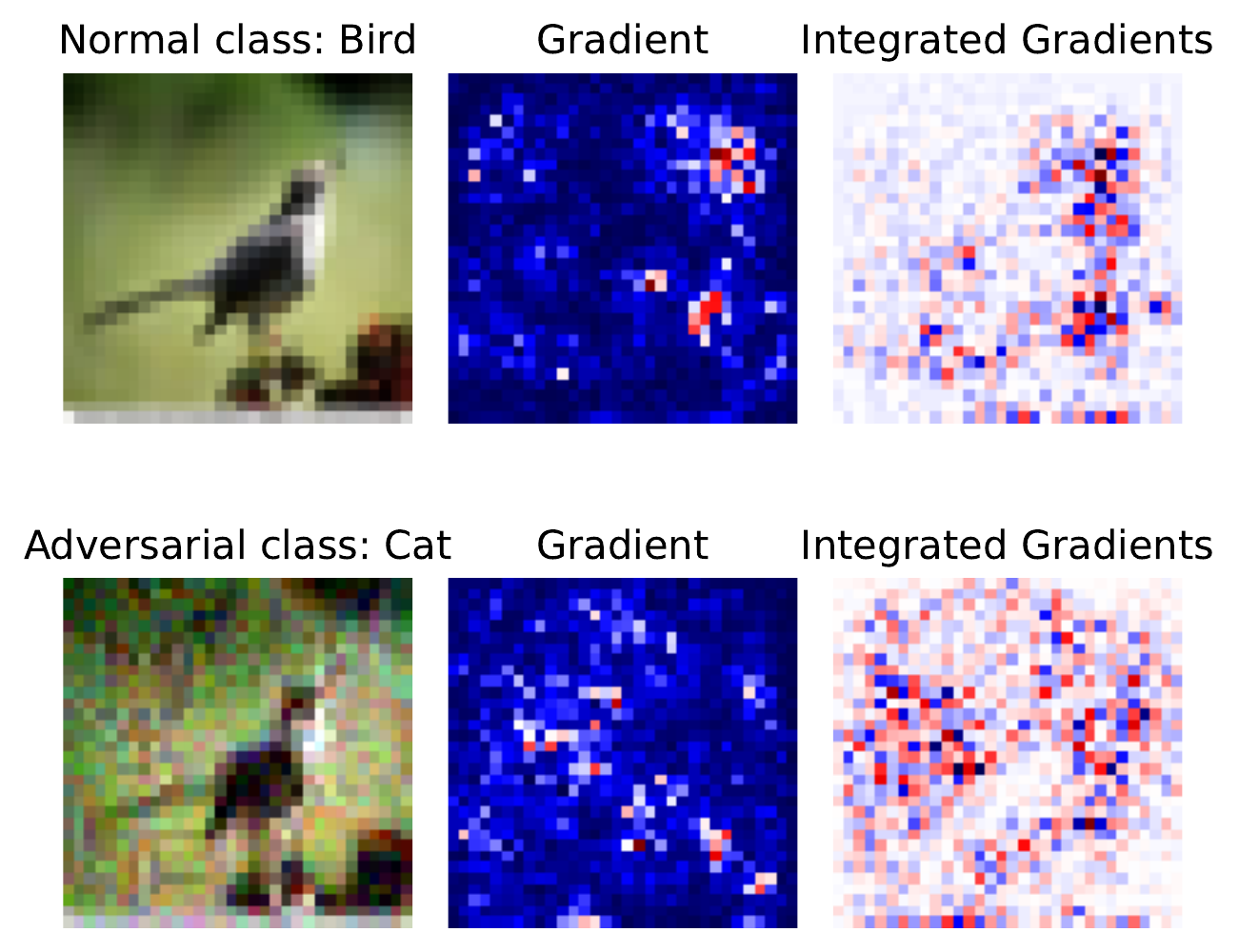} }
    %\caption[caption]{Parsing of unstructured log into structured sequence \footnote{\url{https://github.com/logpai/logparser}}}
    \caption[Caption]{Attribution maps on benign and adversarial CIFAR-10 images for class Bird, with the adversarial class as Cat. Explanation methods used are Gradient and Integrated Gradient. %The first and second row shows heatmaps for benign images and adversarial images.
    }
    \label{fig:cifarplot}
\end{figure}

% \textcolor{blue}{This para should be in related work->}

Deep learning models can be made to produce false predictions by making imperceptible perturbations to the model inputs \cite{szegedy2013intriguing,papernot2017practical,goodfellow2014explaining,madry2017towards}.  These are known as adversarial attacks. Two main approaches have been used to build defenses against these attacks: a) modify network training to build more robust models (e.g., adversarial training \cite{goodfellow2014explaining, madry2017towards, kurakin2016adversarial}), and b) detect and remove manipulated examples that can cause a deep learning model to produce incorrect predictions without modifying neural network training \cite{feinman2017detecting, ma2018characterizing}. Recent research shows a strong connection between explainability and adversarial robustness \cite{chalasani2020concise,etmann2019connection}. They have shown that adversarial training promotes concise and stable explanations and that training a model to achieve stable explanations can improve its adversarial robustness. Additionally, some studies have investigated using feature attribution techniques to detect adversarial samples \cite{jha2019attribution,wang2020interpretability,yang2020ml}.

\textit{\ul{Model \& Dataset}}: For the MNIST classifier, we used LeNet \cite{lecun1998gradient}, whereas, for the CIFAR-10 classifier, we use ResNet-50 \cite{he2016deep}. The Cleverhans \cite{papernot2018cleverhans} library generates the adversarial samples from the projected gradient descent (PGD) \cite{madry2017towards} attack for both classifiers. The PGD attack searches for adversarial samples by iteratively modifying the given test sample with perturbations and clipping them within an $\epsilon$ neighborhood. %Showing the output of an image dataset model prediction or explanation is not the same as that of a tabular or textual dataset. Therefore, 

% \begin{figure}[h!]
%     \centering
%     \resizebox{.6\columnwidth}{!}{
%  \includegraphics[width=.4\textwidth]{figures/CIFARStatisticMedianAbsDevPlot.pdf} }
%     %\caption[caption]{Parsing of unstructured log into structured sequence \footnote{\url{https://github.com/logpai/logparser}}}
%     \caption[Caption]{Measure of statistical dispersion for CIFAR-10 images using Gradient method.}
%     \label{fig:cifarplotMedDev}
% \end{figure}

\begin{figure}[h!]
    \centering
    \resizebox{1\columnwidth}{!}{

  \includegraphics[width=.8\textwidth]{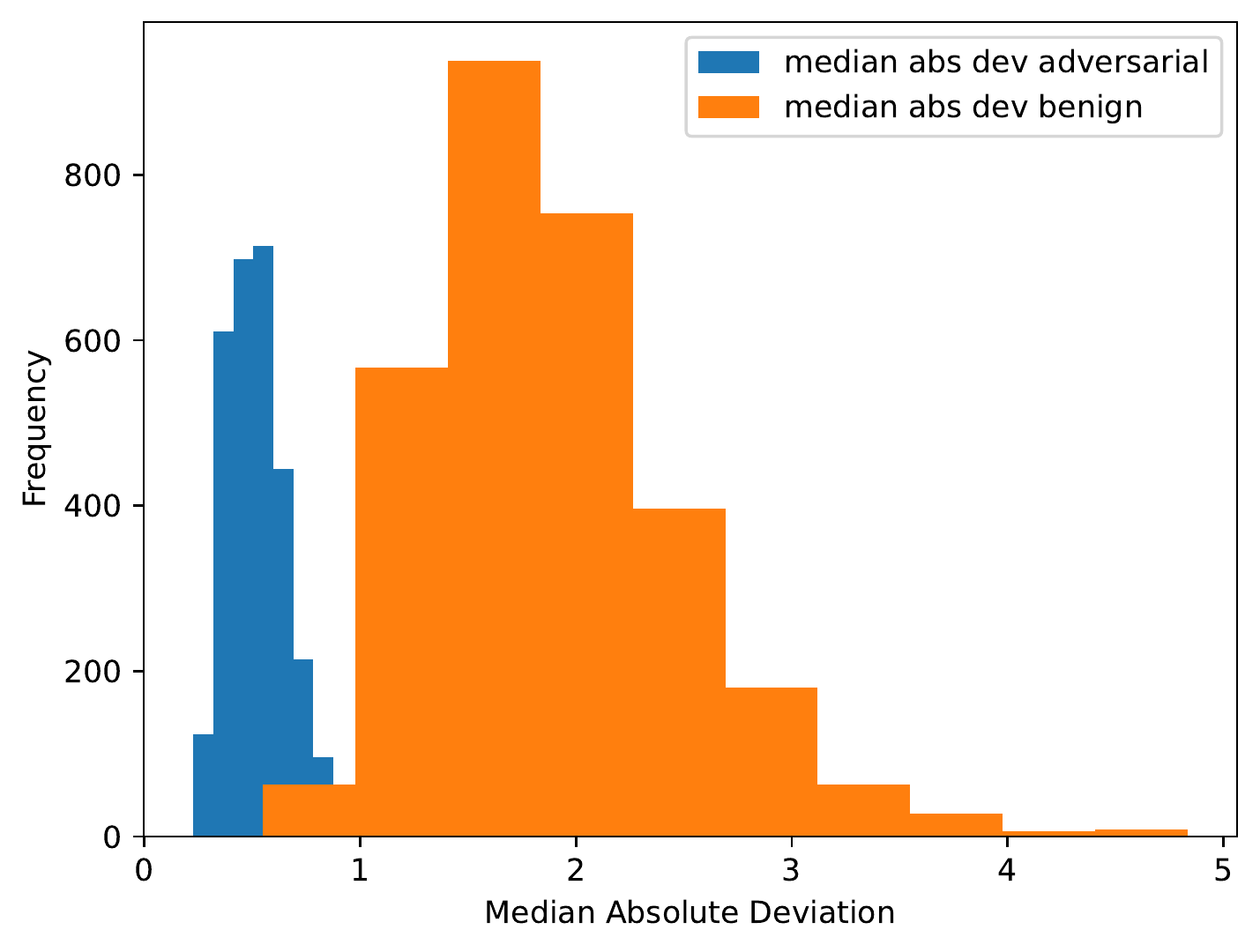}    
 \includegraphics[width=.8\textwidth]{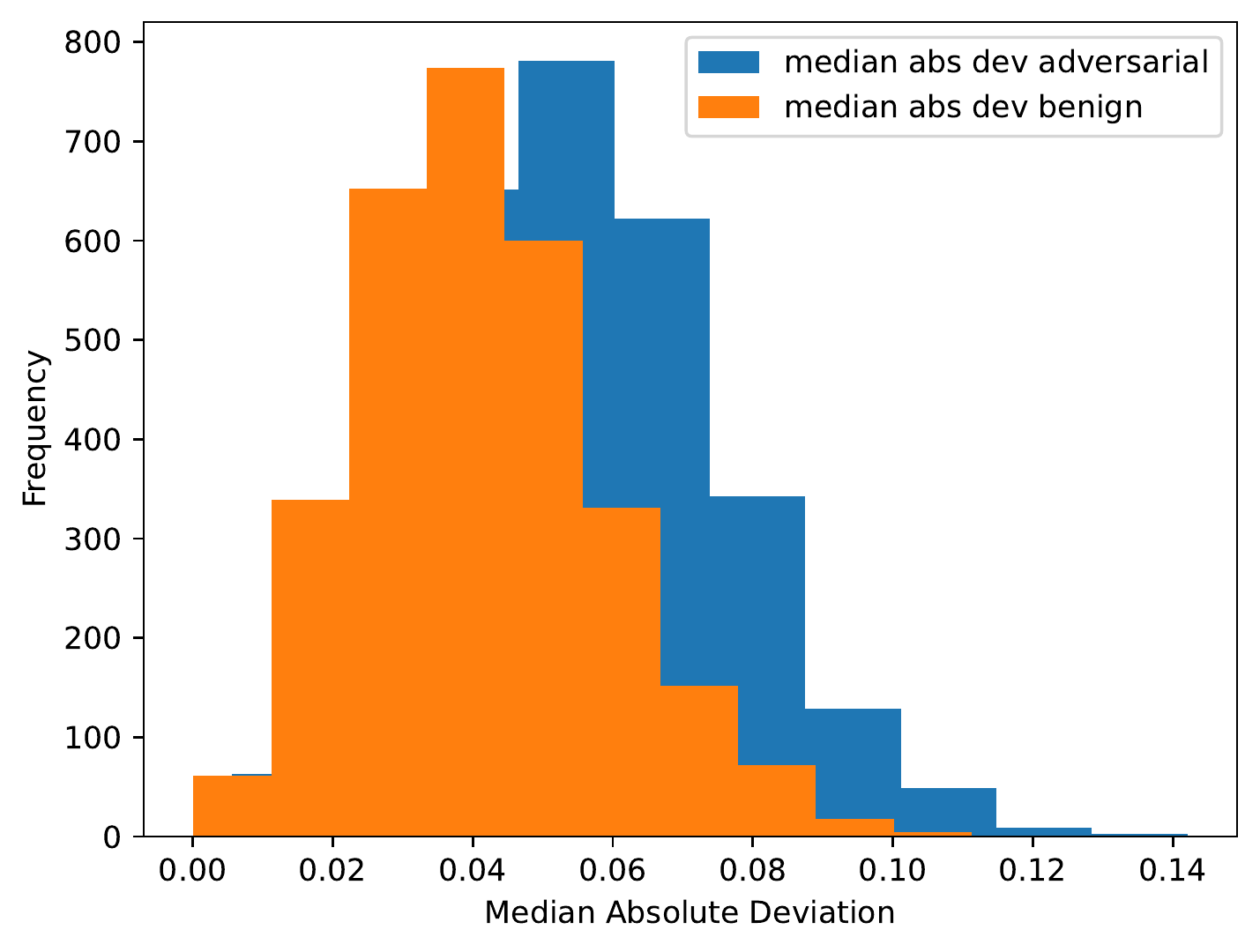}

 }
    %\caption[caption]{Parsing of unstructured log into structured sequence \footnote{\url{https://github.com/logpai/logparser}}}
    \caption[Caption]{Measure of dispersion for CIFAR-10 using (l)Gradient Method, (r)Integrated Gradient. X-axis is Mean Absolute Deviation; Y-axis is Frequency.}
    \label{fig:igcifarplotMedDev}

  %  fig:cifarplotMedDev
\end{figure}

\textit{\ul{Using Explanation Methods}}: Features attributed to adversarial samples have been shown to differ from their benign counterparts due to adversarial perturbations.  This causes model predictions to change, and explanation methods to identify different features as important. Figure \ref{fig:cifarplot} presents heatmaps for corresponding benign and adversarial images of an example bird image in the CIFAR-10 dataset. Due to space limitations, we show the usage of two explainable methods-- gradient and integrated gradient. The heatmaps for the benign and adversarial images are visibly different. Research using the CIFAR-10 and MNIST datasets has shown that differences in feature attribution can be computed using statistics of dispersion, such as median absolute deviation (MAD) (see Figure \ref{fig:igcifarplotMedDev}(l)), which is the median of the absolute deviations from the vector's median. Other statistical measures, such as the interquartile range (IQR), have been used for the same datasets with positive outcomes. Based on this observation, we use a threshold to reject adversarial samples.

% \textit{\ul{Evaluation}}: We apply the gradient explanation method to compute feature attribution for all test instances. Additionally, we use two statistical approaches for identifying adversarial images: median absolute deviation (MAD) and coefficient of inter-quartile range (IQR).

\textit{(a) Quantitative Analysis}: We applied the following threshold strategy based on our observation of the statistical dispersion of feature attribution: 

\begin{enumerate} 

\item For the MNIST dataset, we classified a sample as adversarial if the coefficient of IQR of its feature attribution was less than 0.925; otherwise, we classified it as benign. We used the median absolute deviation to classify a sample as adversarial if the MAD of its feature attribution was greater than 0.011; otherwise, we classified it as benign. 

\item For the CIFAR-10 dataset, we classified a sample as adversarial if the coefficient of IQR of its feature attribution was less than 0.5; otherwise, we classified it as benign. We used the median absolute deviation to classify a sample as adversarial if the MAD of its feature attribution was less than 0.75; otherwise, we classified it as benign. 

\end{enumerate}

\begin{table}[]
\centering
\caption{Performance on adversarial image detection using threshold strategy on feature attribution statistics}
\label{tab:stats}
\resizebox{0.4\textwidth}{!}{%
\begin{tabular}{@{}cll@{}}
\toprule
Dataset/Statistics & Coefficient of IQR                                                           & \multicolumn{1}{c}{Median Absolute Deviation}                               \\ \midrule
MNIST   & \begin{tabular}[c]{@{}l@{}}Precision: 99.38\%\\ Recall: 93.64\%\end{tabular} & \begin{tabular}[c]{@{}l@{}}Precision: 82.89\%\\ Recall: 82.36\%\end{tabular}  \\ \hline
CIFAR-10   & \begin{tabular}[c]{@{}l@{}}Precision: 90.12\%\\ Recall: 82.85\%\end{tabular}   & \begin{tabular}[c]{@{}l@{}}Precision: 99.63\%\\ Recall: 93.50\%\end{tabular} \\ \bottomrule
\end{tabular}%
}
\end{table}

Our experimental results (see Table \ref{tab:stats}) show  that employing an appropriate statistical dispersion measure and threshold strategy specific to each dataset is possible to detect adversarial samples. However, this method relies on explanation methods that produce feature attributions that are statistically distinct for benign and adversarial samples. Unfortunately, the integrated gradient does not provide statistically significant discriminative information on feature attribution for adversarial samples (see Figure \ref{fig:igcifarplotMedDev}(r)). This approach is also not suitable for high-dimensional images like ImageNet and cannot detect mixed confidence attacks like C\&W \cite{carlini2017towards}. Moreover, different datasets necessitate the usage of distinct statistical methods and thresholds.

\textit{(b) Qualitative Analysis}: We interviewed two security experts experienced in adversarial image detection who possess an awareness of explainable methods. The experts were asked to interpret Figure \ref{fig:cifarplot} and share observations from the heatmaps. One expert was critical of the explanations, as they found little difference between the normal and adversarial heatmaps. The other expert observed disparities in the heatmaps between the benign and adversarial samples, with the colored pixels seen as concentrated in the benign heatmap and scattered in the adversarial counterpart. However, both experts were critical of the explanations, with one remarking that they ``need to be validated'' before being used in decision-making. We presented additional information through Figure \ref{fig:igcifarplotMedDev}(l). One reviewer suggested that the new information could help build defenses but cautioned that an adaptive attack could be designed to influence the distribution of feature attribution. Both reviewers inquired whether the dispersion was observable across all image types and using all explanation methods, which was not the case.

% summary table
\begin{table*}[]
\centering
\caption{Comparison of explanation methods. $\uparrow$ and $\downarrow$ indicate where larger and smaller values are better. }
\label{tab:evaluationexplanation}
\resizebox{\textwidth}{!}{%
\begin{tabular}{@{}lccccccccl@{}}
\toprule
\textbf{Method/Metrics} & \multicolumn{1}{l}{\textbf{Faithfulness$\uparrow$}} & \multicolumn{1}{l}{\textbf{Continuity$\downarrow$}} & \multicolumn{1}{l}{\textbf{Monotonicity$\uparrow$}} & \multicolumn{1}{l}{\textbf{Max-Sensitivity$\downarrow$}} & \multicolumn{1}{l}{\textbf{\begin{tabular}[c]{@{}l@{}}Relative \\ Output Stability$\downarrow$\end{tabular}}} & \multicolumn{1}{l}{\begin{tabular}[c]{@{}l@{}}\textbf{Model parameter} \\ \textbf{randomisation}$\uparrow$\end{tabular}} & \textbf{Sparsity$\uparrow$} & \textbf{Complexity$\downarrow$} & \textbf{Rating} \\ \midrule
\textbf{Gradient} & 0.105 & 12.339 & 0.139 & 0.726 & 6.840 & 0.804 & 0.443 & 3.731 &  $\star$$\star$$\star$\\
\textbf{GradentXInput} & 0.668 & 6.067 & 0.271 & 0.315 & 12.732 & 0.993 & 0.874 & 2.217 & $\star$$\star$$\star$$\star$ \\
\textbf{Integrated Gradient} & 0.777 & 4.868 & 0.271 & 0.183 & 16.434 & 0.994 & 0.875 & 2.207 & $\star$$\star$$\star$$\star$$\star$ \\
\textbf{DeepLift} & 0.667 & 4.948 & 0.271 & 0.242 & 16.435 & 0.513 & 0.873 & 2.218 &  $\star$$\star$$\star$$\star$\\
\textbf{GradientShap} & 0.746 & 5.898 & 0.271 & 0.262 & 16.798 & 0.497 & 0.875 & 2.201 & $\star$$\star$$\star$$\star$ \\
\textbf{LIME} & 0.217 & 14.665 & 0.002 & 0.249 & 12.148 & 0.134 & 0.562 & 3.473 &  $\star$$\star$$\star$\\
\textbf{SHAP} & 0.171 & 8.747 & 0.012 & 0.390 & 12.276 & 0.583 & 0.426 & 3.759 & $\star$$\star$ \\
\textbf{Occlusion} & 0.676 & 4.867 & 0.276 & 0.408 & 10.527 & 0.962 & 0.743 & 2.712 & $\star$ \\ \bottomrule
\end{tabular}%
}
\end{table*}

\subsection{Analysis}\label{sec:evaluationsummary}
%An explanation method that achieves good results both qualitatively and quantitatively are reliable and dependable for real-life tasks. 
A trustworthy explanation method should achieve better results in quantitative and qualitative evaluation. For quantitative evaluation, we computed the average performance of all methods and summarized the results in Table \ref{tab:evaluationexplanation}, assigning an overall rating to each method. Gradient-based methods consistently outperformed perturbation methods (LIME, SHAP, Occlusion). 

Among the gradient methods, Integrated Gradient achieved the highest scores in faithfulness, max-sensitivity, model parameter randomization, and sparsity. Additionally, Integrated Gradient was preferred over other explanation methods in the qualitative analysis of log anomaly detection and malware detection. Occlusion provided poor explanations qualitatively but achieved good scores in faithfulness, sparsity, and model parameter randomization, quantitatively (see Table \ref{tab:evaluationexplanation}). %Across all use cases in Table \ref{tab:pdfUseCaseExplanations}, the usability of the explanation methods was not rated positively due to the low quality, low coherence, and low actionability scores. 

Expert adaptability to explanation methods was surprisingly positive.  We initially expected the security experts to require extensive guidance or prior knowledge of explainable methods.  However,  during the interviews, they were able to analyze the explanations after providing a short introduction to explanations. All experts acknowledged that the current explainability methods lack the ability to provide a meaningful explanation for real-world tasks. Consequently, our study shows that from the usability perspective of security end-users, the current explanation methods do not meet user requirements. A summary of the qualitative analysis is provided in Table \ref{tab:qualeval}.
\section{Concerns with explanations}\label{sec:concerns}
% \textcolor{blue}{we can remove this para. since we repeat it on 6.1 and 6.2}
% Security systems are complex, involving multiple stakeholders and intricate system models \cite{vigano2020explainable}. Model explanations can be misused to reconstruct models and training data \cite{shokri2021privacy}, evade detection of the model \cite{demetrio2019explaining}, and poison the training set \cite{kuppa2021adversarial}, thereby jeopardizing the privacy of users and the integrity of a security system. The multifaceted nature of security systems makes explainable security a challenging endeavor. We identify the following major concerns:

\subsection{Security concerns}
% \textcolor{blue}{1.1)} The requirements of an explanation method for a system designer and security analyst in a security operation center are widely different. While the system designer (maintainer) seeks to ensure that the system is working as intended and improve the model performance based on feedback from several test cases, a SOC analyst only seeks reliable information to validate threats quickly and efficiently \cite{nadeem2022sok}. Explanation results produced by the explanator for the system designer and analyst should be different to suit their needs.
 
%\textcolor{blue}{1.2)} A security system is often vulnerable to adversary attacks. An adversary compromises a system's confidentiality, integrity, and availability by launching different attacks. Model explanations can be misused for reconstructing model and training data \cite{shokri2021privacy}, evade detection of the model \cite{demetrio2019explaining}, and poison the training set \cite{kuppa2021adversarial}, thus compromising the privacy of users and integrity of a security system. Such multifaceted nature of security systems makes explainable security a challenging endeavor.

\textbf{Error tolerance:} In non-security applications, errors in explanations, such as including insignificant details like a few pixels in an image, are usually tolerable as they do not have significant consequences. Partially correct explanations can be sufficient for an intuitive understanding. However, in security explanations, there is no room for error \cite{guo2018lemna}. Security applications require high-quality and robust explanations. Even a tiny mistake, such as a single byte of code in binary analysis or one log sequence in system logs, can result in severe misunderstandings. For instance, in Section \ref{sec:usecase}, a SOC analyst relies on an explanation method to evaluate an alert to diagnose a threat to confirm or dismiss the alert. However, an incorrect explanation or alert validation can affect the organization's cyber infrastructure. Therefore, an explanation method must ensure that it provides accurate impressions to an analyst.

\par 

\textbf{Evaluation of explanation:} Explanations are evaluated based on their quality and usefulness. Quality is measured using quantitative metrics (discussed in Section \ref{sec:evaluation}). Although several metrics are proposed in the literature, there is a lack of uniformity in their acceptance and evaluation \cite{warnecke2020evaluating}. Additionally, relying solely on quantitative metrics can lead to insufficient and inaccurate evaluations. As mentioned in Section \ref{sec:evaluationsummary}, an explanation method that produces inaccurate explanations may still receive high scores, which can mislead an end-user about its performance. The usefulness of an explanation is determined by its benefits to an end-user. Therefore, qualitative evaluation can complement quantitative evaluation and improve the comprehension of explanations, enhancing its usability. As discussed in Section \ref{sec:evaluation}, explanation methods should also be evaluated based on usability criteria such as coherence and actionability. %In images and video, a qualitative explanation using saliency highlighting \cite{selvaraju2017grad} is often used as it is a basic yet intuitive method of understanding how a model is making certain predictions. However, for security applications with varying type of dataset, there is no single best qualitative approach, such as heatmap or rules. %Explanations generated for security should be clear and easily understandable.

%proposed pipeline
\begin{figure*}
    \centering
  \includegraphics[width=0.7\textwidth]{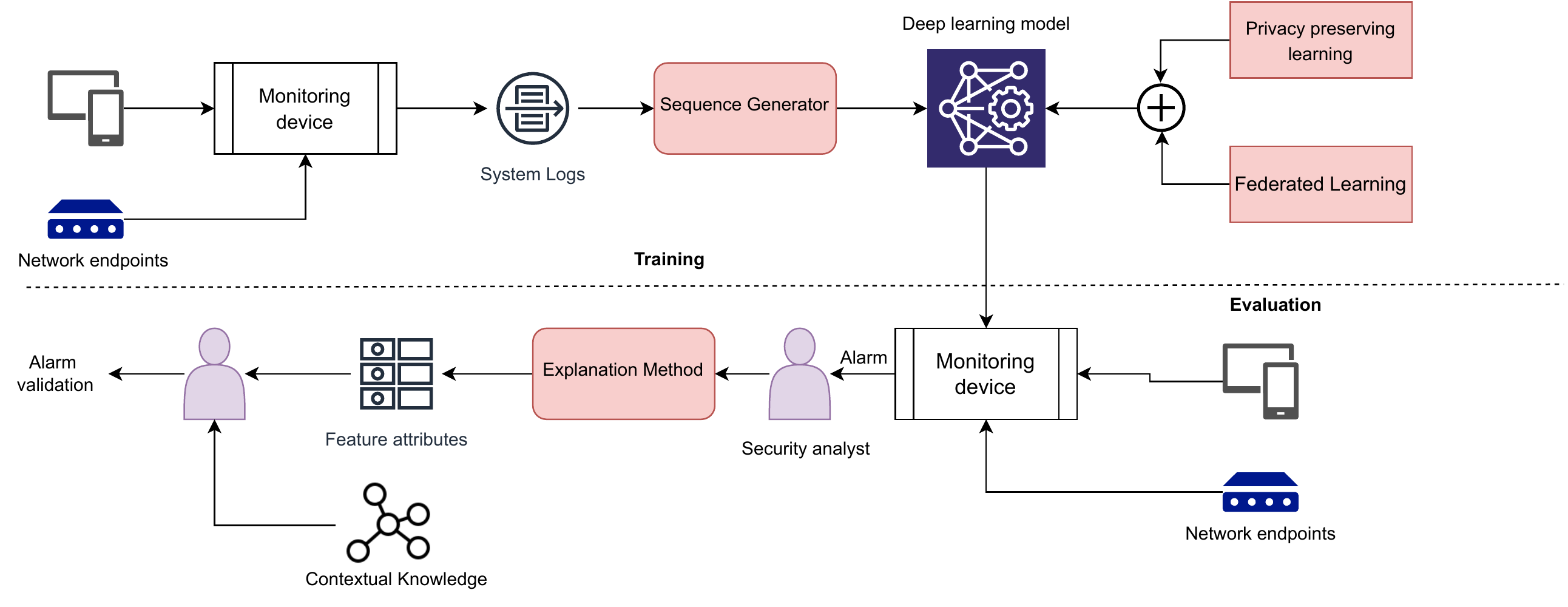} 
    %\caption[caption]{Parsing of unstructured log into structured sequence \footnote{\url{https://github.com/logpai/logparser}}}
    \caption[Caption]{Proposed pipeline for accuracy, privacy, and interoperability. To preserve privacy, deep learning model is trained with additional privacy-preserving constraints. Alerts are validated during evaluation using feature attributes from an explanation method and additional contextual knowledge.}

        % \caption[Caption]{Proposed pipeline for accuracy, privacy, and interpretability. During the training phase, the deep learning model for system anomaly detection is trained with additional privacy-preserving constraints that preserve model privacy. A security analyst validates alarms in the evaluation phase using feature attributes from an explanation method and additional contextual knowledge provided to them.}
        
    \label{fig:pipeline}
\end{figure*}

\subsection{Privacy concerns}
%Explainable machine learning is used by end-users for decision support or model verification. However,....  %A model explanation can reveal sensitive information about the training data, undermining data privacy \cite{zhou2020transparency}. 
An adversary can exploit model explanations to strengthen their attack, potentially compromising the integrity and confidentiality of a model. For instance, an adversary can use model explanations to obtain sensitive information from a dataset using inference attacks \cite{Shokri2017MembershipIA}. As demonstrated in \cite{shokri2021privacy}, an adversary can reconstruct a significant portion of a dataset using model explanation methods. Gradient-based explanations perform better than perturbation-based methods but still disclose important training data information. Studies \cite{dombrowski2019explanations, heo2019fooling, zhang2020interpretable} have also shown that explanations can be manipulated to produce targeted explanations, thus deceiving an end-user about a model's performance. Therefore, it is essential to be aware of any explanation method risks and to design model explanations that protect both data and models \cite{patel2022model}.
%\hspace{-0.5cm}
\subsection{End-user concerns}

Explanation methods play a crucial role in security applications, forming an essential part of security analysts' workflow. However, integrating explanation solutions into existing security tools can be challenging, as they must be usable and computationally efficient \cite{nyre2021considerations}. A study shows that analysts are still hesitant to use explanation methods due to the additional time for investigation and the complex nature of explanations \cite{crowley2019common}. Therefore, it is essential to evaluate the intuitiveness, coherence, and usability of explanations for different datasets and explanation methods \cite{afzaliseresht2022explainable, bertrand2022cognitive}.

\textbf{Comprehensibility:}
The comprehensibility of explanations can be affected by the type of explanation method or the dataset. While feature attribution-based methods are suitable for all datasets and models, they provide feature relevance scores that require domain expertise to understand \cite{alahmadi202299}. For example, when we analyze the results of Section \ref{sec:usecaseI} and \ref{sec:usescaseII}, the explanations are less intuitive given that there is no detailed (or short) verbal explanation that led to such detection. Rule-based explanations, such as ANCHOR, \cite{ribeiro2018anchors} can improve explanation comprehension if they support both the model and dataset. Nevertheless, any explanation method employed in security applications must provide accurate and reliable information that can be both easily and rapidly understood and does not require additional thought or processing by the end user.

\section{Challenges \& research directions}\label{sec:challenges}

\subsection{Transformer Models \& Explanations}

%Transformer models have achieved state-of-the-art results in natural language sequence tasks due to their ability to capture long-term dependencies in language, making them suitable for sequence prediction in security. However, it is essential to ensure that complex transformer models are transparent to understand the model predictions. Since they rely heavily on skip connections and attention operators, one must exercise caution when applying explanation methods to transformer models. For instance, in transformer models, the attention mechanism \cite{bahdanau2014neural} assigns weights to tokenized words in a sequence and can be used to understand sequence prediction. However, such an approach lacks faithfulness and stability \cite{burstein2019proceedings}. Alternatively, one can employ post-hoc explanation methods like hidden state evolution and neuron activations \cite{alammar2021ecco} to understand transformer predictions. Nonetheless, transformer models are complex and tend to be computationally slow, requiring careful evaluation for practical deployment in security.

Transformer models are widely used in natural language sequence tasks due to their ability to capture long-term dependencies in language resulting in state-of-the-art results \cite{vaswani2017attention}. This property makes them suitable for sequence prediction in security. However, ensuring transparency in these models is crucial to understand model predictions. Although explanation methods can be applied to transformer models, caution must be exercised due to their heavy reliance on skip connections and attention operators. For instance, the attention mechanism assigns weights to tokenized words in a sequence, but this approach lacks faithfulness and stability \cite{burstein2019proceedings}. Post-hoc explanation methods like hidden state evolution and neuron activation can be used to understand transformer predictions \cite{alammar2021ecco}. However, transformer models are complex and computationally slow, requiring careful evaluation before practical deployment in security applications.

\subsection{Stable, reliable and robust explanations}

%The lack of reliability, stability, and usability in existing explanation methods has hindered their wide acceptance in security applications \cite{gan2022your, nyre2021considerations}. These methods are vulnerable to different kinds of adversarial attacks as they can be manipulated to produce false explanations or be exploited to leak training data \cite{heo2019fooling, shokri2021privacy}. Such properties inhibit end-users from employing explanation methods even when they are needed to understand black box models. The design of explanation methods must ensure that they produce stable and reliable explanations that are comprehensible to end-users. One research direction is to develop certifiably robust explanation methods that not only generate reliable explanations but also provide certified robustness guarantees for the explanations \cite{huai2022towards}. A challenging but contrasting task is to design an inherently interpretable model with the same performance as complex deep learning models \cite{rudin2019stop}.

The lack of reliability, stability and usability in existing explanation methods has hindered their acceptance in security applications \cite{gan2022your, nyre2021considerations}. For example, adversarial attacks can manipulate these methods to provide false explanations, inhibiting end-user adoption \cite{heo2019fooling, shokri2021privacy}. Therefore, a research direction that has been explored is to develop certifiably robust explanation methods that provide certified robustness guarantees for the explanations \cite{huai2022towards}. A challenging but contrasting task is to design inherently interpretable models that perform as well as complex deep learning models \cite{rudin2019stop}.

\subsection{Addressing accuracy, privacy, and trust}

% \textcolor{blue}{Instead of long section, i have summarized this into 2 para.}

The trustworthiness of black box machine learning models depends on their accuracy, privacy, and explainability \cite{harder2020interpretable}. However, improving one of these properties can compromise others, as improving interpretability with model explanation can compromise privacy, and enhancing accuracy with complex models can hinder model understanding \cite{zhou2020transparency}. We propose a pipeline for a system log anomaly detector to address this trade-off, as shown in Figure \ref{fig:pipeline}. Our proposed pipeline includes a deep learning model for anomaly detection and post-hoc explanation methods to provide model prediction understanding. Additionally, incorporating contextual knowledge with knowledge graphs \cite{jia2018pattern}, we believe alert validation and explanation understanding can be significantly improved as demonstrated in previous works \cite{kruegel2004alert}.
\par
To address privacy concerns in explanation methods, we can adopt privacy-preserving techniques like differentially private model training or federated learning \cite{abadi2016deep, dwork2018privacy, datta2016algorithmic, patel2022model, mcmahan2017communication, parra2022interpretable}. Differentially private model training can ensure that an adversary cannot exploit black-box model explanations to leak information about the training data \cite{datta2016algorithmic, patel2022model}. Federated learning also provides a privacy-preserving solution due to its distributed nature \cite{mcmahan2017communication} and has been used in federated log learning for threat forensics \cite{parra2022interpretable}. Our proposed pipeline can address the trade-off between accuracy, privacy, and explainability by incorporating these privacy-preserving techniques.

\section{Conclusion}

This research paper provides insights into the explainability of deep learning models for security applications. We study the importance of transparency and interpretability of deep learning model predictions by evaluating several post-hoc explanation methods against three security applications. Quantitative analysis reveals that back-propagation-based explanation methods (e.g., Integrated Gradient) are effective in practical settings. However, the qualitative analysis also shows issues related to the actionability and usability of explanation methods in real-world applications. Overall, it is necessary to develop explanation methods that are reliable, robust, and user-friendly to enhance the practical application of deep learning models in real-world security settings. Finally, we propose a novel pipeline to address the trade-offs between privacy, accuracy, and explainability, thereby enhancing the utility of explainable methods in real-world security applications.

\clearpage
\newpage
\bibliographystyle{ACM-Reference-Format}
\bibliography{sample-base}

\appendix

\end{document}